\def\lex{\mathrm{map}_{i\leftarrow \bm p}}
\def\({\left (}
\def\){\right )}

\documentclass[runningheads]{llncs}
\usepackage{caption}
\usepackage{amsmath}
\usepackage{amssymb}
\usepackage{graphicx}
\usepackage[dvipsnames]{xcolor}
\usepackage{hyperref}
\usepackage{booktabs}
\usepackage{floatrow}
\usepackage{bm}
\usepackage{multirow}
\usepackage{caption}

\begin{document}

\title{Sparse Matrix-Based HPC Tomography}

%
%
\author{Stefano Marchesini\inst{1}
\and
Anuradha Trivedi\inst{2}\and
Pablo Enfedaque\inst{3} \and
Talita Perciano\inst{3}
\and
Dilworth Parkinson\inst{4}
}
\authorrunning{S. Marchesini, A. Trivedi, P. Enfedaque, T. Perciano, D. Parkinson}
%
\institute{Sigray, Inc., 5750 Imhoff Drive, Ste I, Concord, CA 94520 USA
\email{smarchesini@sigray.com}
\url{http://sigray.com} \and
Virginia Polytechnic Institute and State University, Blacksburg, VA 24061 USA\\
 \and Computational Research Division, Lawrence Berkeley National Laboratory, 1 Cyclotron Rd. Berkeley, CA  94720 USA \\ 
 \and Advanced Light Source, Lawrence Berkeley National Laboratory, 1 Cyclotron Rd. Berkeley, CA 94720 USA 
}

\maketitle              
\begin{abstract}
 Tomographic imaging has benefited from advances in X-ray sources, detectors and optics to enable novel observations in science, engineering and medicine. These advances have come with a dramatic increase of input data in the form of faster frame rates, larger fields of view or higher resolution, so high performance solutions are currently widely used for analysis.  Tomographic instruments can vary significantly from one to another, including the hardware employed for reconstruction: from single CPU workstations to large scale hybrid CPU/GPU supercomputers. Flexibility on the software interfaces and reconstruction engines are also highly valued to allow for easy development and prototyping. This paper presents a novel software framework for tomographic analysis that tackles all aforementioned requirements. The proposed solution capitalizes on the increased performance of sparse matrix-vector multiplication and exploits multi-CPU and GPU reconstruction over MPI. The solution is implemented in Python and relies on CuPy for fast GPU operators and CUDA kernel integration, and on SciPy for CPU sparse matrix computation. As opposed to previous tomography solutions that are tailor-made for specific use cases or hardware, the proposed software is designed to provide flexible, portable and high-performance operators that can be used for continuous integration at different production environments, but also for prototyping new experimental settings or for algorithmic development. The experimental results demonstrate how our implementation can even outperform state-of-the-art software packages used at advanced X-ray sources worldwide. 
 
\end{abstract}
\section{Introduction}

Ever since Wilhelm Röntgen shocked the world with a ghostly photograph of his wife’s hand in 1896, the imaging power of X-rays has been exploited to help see the unseen. Their penetrating power allows us to view the internal structure of many objects. Because of this, X-ray sources are widely used in multiple imaging and microscopy experiments, e.g. in Computed Tomography (CT), or simply tomography. A tomography experiment measures a transmission absorption image (called radiograph) of a sample  at multiple rotation angles. From 2D absorption images we can reconstruct a stack of slices (\textit{tomos} in Greek) perpendicular to the radiographs measured,  containing the 3D volumetric structure of the sample.
Tomography is used in a variety of fields such as medical imaging, semiconductor technology, biology and materials science.
Modern tomography instruments using synchrotron-based light sources can achieve measurement speeds of over 200 volumes per second using 40 kHz frame rate detectors  \cite{garcia2019}. Tomography can also be combined with microscopy techniques to achieve resolutions down to  a single atom using electrons \cite{scott2012electron}. Its experimental versatility has also been exploited by combining it with spectroscopic techniques, to provide chemical, magnetic or even atomic orbital information about the sample.

Nowadays, tomographic analysis software faces three main challenges. 1) The volume of the data is constantly increasing as X-ray sources become brighter and newer generation detectors increase their resolution and acquisition frame rate. 2) Instruments from different facilities (or even from the same one) present a variety of experimental settings that can be exclusive to said instrument, such as the geometry of the measurements, the data layout and format, noise levels, etc. 3) New experimental use cases and algorithms are frequently explored and tested to accommodate new science requisites. These three requirements strongly force tomography analysis software to be HPC and flexible, both in terms of modularity and interfaces, as well as in hardware portability. Currently, TomoPy \cite{Tomopy_Gursoy} and ASTRA \cite{ASTRA_Aarle} are the most popular solutions for tomographic reconstruction at multiple synchrotron and tabletop instruments.  TomoPy is a Python-based open source framework optimized for performance using a C backend that can process a variety of data formats and algorithms. ASTRA is a tomography toolbox accelerated using both GPU and CPU computing and it is also available through TomoPy \cite{pelt2016integration}. Although both solutions are highly optimized at different levels, they do not provide the level of flexibility required to be easily extendable by third parties regarding solver modifications or accessing specific operators. 

In this work we present a novel framework that focuses on providing multi-CPU and GPU acceleration with flexible operators and interfaces for both 1-step and iterative tomography reconstruction. The solution is based on Python 3 and relies on CuPy, mpi4py, SciPy and NumPy to provide transparent CPU/GPU computing and innocuous multiprocessing through MPI. The idea is to provide easy HPC support without compromising the solution lightweight so that development, integration and deployment is streamlined. The current operators are based on sparse matrix-vector multiplication (SpMV) computation which benefit from preexisting fast implementations on both CuPy and SciPy and provide faster reconstruction time than direct dense computation~\cite{jackson1991selection}. By minimizing code complexity, we can efficiently implement advanced iterative techniques~\cite{venkatakrishnan2013plug,mohan2014model} that are not normally implemented for production, also due to their computational complexity; prior implementations could take up to a full day of a supercomputer to reconstruct a single tomogram \cite{venkatakrishnan2016making}. The high level technologies and modular design employed in this project permits the proposed solution to be particularly flexible, both for exploratory uses (algorithm development or new experimental settings), and also in terms of hardware: we can scale the reconstruction from a single CPU, to a workstation using multiple CPU and GPU processors, to large distributed memory systems. The experimental results demonstrate how the proposed solution can reconstruct datasets of 68 GB in less than $5$ seconds, even surpassing the performance of TomoPy's fastest reconstruction engine by $2.2$X. This project is open source and available at~\cite{xpack}.



The paper is structured as follows: Section 2 overviews the main concepts regarding tomography reconstruction. Section 3 presents the proposed  implementation with a detailed description of the challenges behind its design and the techniques employed, and Section 4 assesses its performance through experimental results. The last section summarizes this work.

 






\section{Tomography}

\begin{figure}[ht!]
    \centering
    \includegraphics[width=0.9\textwidth]{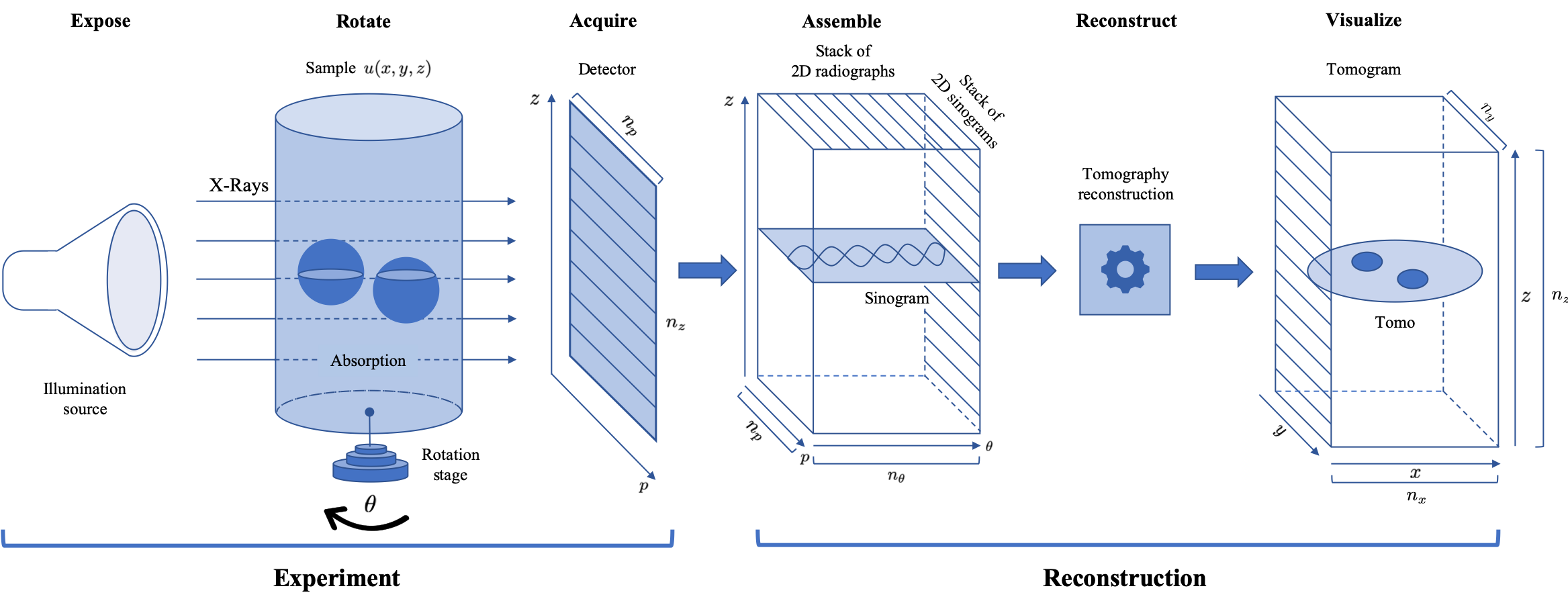}
    \caption{Overview of a tomography experiment and  reconstruction. A 3D sample is  rotated at angles $\theta=0, \dots, 180^{\circ}$ as X-rays produce  2D radiographs onto the detector.  The collection of radiographs  is combined to provide a sinogram for each detector row. Each sinogram is then processed to generate a 2D reconstructed slice, the entire collection of which can be assembled into a 3D tomogram. 
    \label{fig:tomography}\vspace{-0.5cm}}
\end{figure}

Tomography is an imaging technique based on measuring a series of 2D radiographs of an object rotated at different angles relative to the direction of an X-ray beam~(Fig. \ref{fig:tomography}). A radiograph of an object at a given angle is made up of line integrals (or projections). The collection of  projections from different angles at the same slice of the object is called sinogram (2D); and the final reconstructed volume is called tomogram (3D), which is generally assembled from the independent reconstruction of each measured sinogram.
\vspace{-0.2cm}

Physically, the collected data measures attenuation, which is the loss of flux through a medium. 
When the X-ray beams are parallel along the optical axis, the beam intensity impinging on the detector is given by:
\[
    I_\theta (p,z)= I_0 e^{ - \mathcal{P}_{\theta}(p,z)},\quad
    \mathcal{P}_\theta(p,z) = \int u\left(p\cos \theta-s\sin \theta, p\sin\theta + s\cos\theta,z\right)ds,
\]
where $u(x, y,z)$ is the attenuation coefficient as a function of position  $\bm x=(x,y,z)$  in the sample, $I_0$ is the input intensity collected without a sample,
$\mathcal{P}_\theta$ is the projection  after rotating $\theta$ around the $z$ axis, and $(p,z)$ are the coordinates on the detector that sample the data onto $(n_p \times n_z)$ detector pixels.
The negative log of the normalized data provides the projection, also known as X-ray transform:
    \[\mathcal{P}_\theta(p,z) = -\ln\left(\frac{I_\theta(p,z)}{I_0(p,z)}\right), 
    \]
    with element-wise log and division.
    The Radon transform (by H.A. Lorentz \cite{bockwinkel1906propagation})  at a fixed $z$ is then given by the set of $n_\theta$ projections for a series of angles $\theta$:
    \[
    \mathrm{Radon}_{(\theta,p)\leftarrow(x,y)} u(x,y,z)= \mathrm{Sinogram}_z(p,\theta)=\mathcal{P}_\theta (p,z).
    \]
 \subsection{Iterative Reconstruction Techniques\label{sec:irt}}
 
 The tomography inverse problem can be expressed as follows:
\[
\textrm{To find $u$ s.t. } \mathrm{Radon}(u)=-\log(I/I_0).
\]
The pseudo-inverse $\mathrm{iRadon}(-\log(I/I_0))$ described below provides the fastest solution to this problem, and it is typically known as \emph{Filtered Back Projection} in the literature. When implemented in Fourier space, the algorithm is referred to as \emph{non-uniform inverse FFT} or \emph{gridrec}.

The inverse problem can be under-determined and ill-conditioned when the number of angles is small. The equivalent least squares problem is:
\[
\arg\min_u \|\mathbb{P} \left(\mathrm{Radon}(u) + \log(I/I_0) \right)\|,
\]
where $\mathbb{P}=\mathcal{F}^\dagger\mathcal{D}^{1/2} \mathcal{F}$ is a preconditioning matrix, with $\mathcal{D}$ a diagonal matrix, and $\mathcal{F}$ denotes a 1D Fourier transform. Note that $\mathcal{F}$ does not need to be computed when  using the Fubini-Radon operator (see below).
The model-based problem is:
\[
\arg\min_u \|\widehat{\mathbb{P}} \left(\mathrm{Radon}(u)+\log(I/I_0) \right)\|_{w}+\mu \cdot \mathrm{Reg}(u),
\]
where $\|\cdot\|_w$ is a weighted norm to account for the noise model, $\widehat{\mathbb{P}}$ may incorporate streak noise removal \cite{maia2010compressive} as well as preconditioning,  
$\mathrm{Reg}$ is a regularization term such as the Total Variation norm to account for prior knowledge about the sample, and $\mu$ is a scalar parameter to balance the noise and prior models. 

Many algorithms have been proposed over the years including Filtered Back Projection (FBP),  Simultaneous Iterative Reconstruction Technique (SIRT), Conjugate Gradient Least Squares (CGLS), and Total Variation (TV) \cite{bouman1993generalized}. FBP can be viewed as the first step of the preconditioned steepest descent when starting from 0.
To solve any of these problems, one needs to compute a Radon transform and its (preconditioned) adjoint operation multiple   times. 
\subsection{The Fubini-Radon transform}    

    \begin{figure}[tb] 
        \centering
        \includegraphics[width=0.9\textwidth]{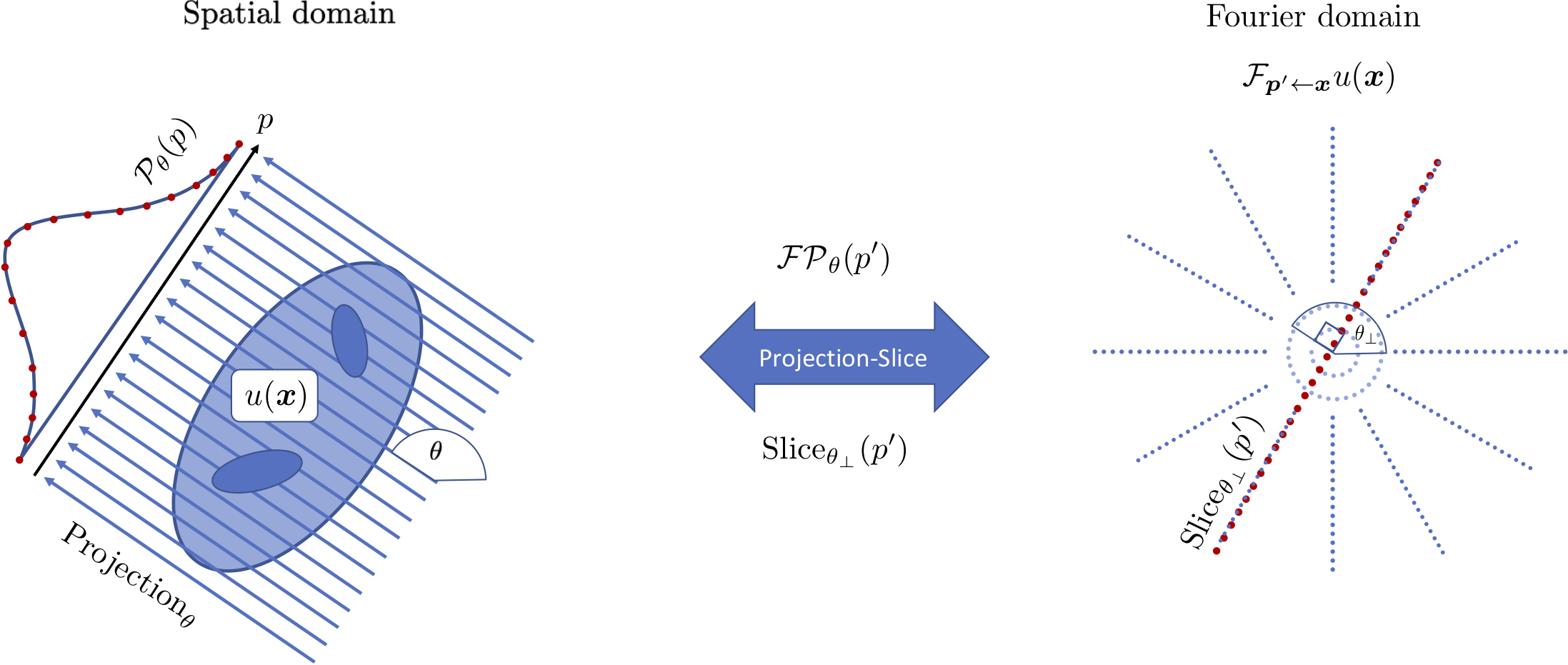}
        \caption{ 
        Depiction of the Fubini-Radon transform (\ref{eq:fourier-slice}), based on the Fourier slice theorem. The projection  $\mathcal{P}_\theta(p,z)$ at a given angle $\theta$, height $z$,  is related to an orthogonal 1D slice (different from the tomographic slice) by a Fourier transform: $\mathcal{F}_{p^\prime\leftarrow p}\left(\mathcal{P}_{\theta}(u(\bm x))\right)=\text{Slice}_{\theta_\perp,p^\prime } \mathcal{F}_{\bm p^\prime \leftarrow \bm x}(u(\bm x))$. The slicing interpolation  between the Cartesian and polar grid is the key step in this procedure and can be implemented with a sparse matrix operation.   \vspace{-0.4cm}}
        \label{fig:fourier_slice}
    \end{figure}

One of the most efficient ways to perform Radon($u$) 
is to use the Fourier central slice theorem by Fubini \cite{grunbaum}. It consists on performing first a 2D Fourier transform, denoted by $\mathcal{F}$ of $u$, and then interpolating the transform onto a polar grid, to finally 1D inverse Fourier transforming the points along the radial lines:
\begin{align}
\text{Radon}_{(\theta,p)\leftarrow(x,y)} =
\mathcal{F}^\ast_{p\leftarrow p^\prime} 
\text{Slice}_{(\theta_\perp,p^\prime)\leftarrow (p_x^\prime,p_y^\prime)}\, \mathcal{F}_{(p_x^\prime,p_y^\prime)\leftarrow(x,y)} 
\label{eq:fourier-slice}
\end{align}
We will refer to this approach as the \emph{Fubini-Radon} transform~(Fig. \ref{fig:fourier_slice}). Numerically, the slicing interpolation between the Cartesian and polar grid in the Fubini-Radon transform is the key step in the procedure. It can be carried out using a \emph{gridding} algorithm that maintains the desired accuracy with low computational complexity. The \emph{gridding} algorithm essentially allows us to perform a \emph{non-uniform FFT}. The projection operations require $\mathcal{O}(n^2  )\cdot n_\theta$ arithmetic operations when computed directly using  $n=n_p=n_x=n_y$ discretization of the line integrals, while the Fubini-Radon version requires $\mathcal{O}(n)\cdot n_\theta$+$\mathcal{O}(n^2\log(n))$ operations, where the first term is due to the slicing operation and the second term is due to the two dimensional FFT. For sufficiently large $n_\theta$, the Fubini-Radon transform requires fewer arithmetic operations than the standard Radon transform using projections. 
Early implementations on GPUs used ad hoc kernels to deal with atomic operations and load-balancing of the highly non-uniform distribution of the polar sampling points~\cite{maia2010compressive}, but became obsolete with new compute architectures. In this work we implement the slicing interpolation using a sparse matrix-vector multiplication. The SpMV and SpMM operations are level 2 and level 3 BLAS functions which have been heavily optimized (see e.g. \cite{yelik}) on numerous architectures for both CPUs and GPUs.

 \section{Radon Transform by Sparse Matrix Multiplication}
The gridding operation requires the convolution between regular samples and a kernel to be calculated at irregular sample positions, and vice versa for the inverse gridding operation. To maintain high numerical accuracy and minimize the number of arithmetic operations, we want to limit the width of the convolution kernel. Small kernel width can be achieved by exploiting the finite sample dimensions ($u(x)>0$ in the field of view) using a pair of functions $k^\star(\bm x),  k(\bm x)$  so that $k^\star(\bm x) k(\bm x) =\{ 1$  if $u(\bm x)>0$,  0 otherwise$\}$. By the convolution theorem:
\[
u=( k^\star\circ k) \circ u =k^\star \circ \mathcal{F}^{-1} (K \circledast \mathcal{F} u ),
\; K=\mathcal{F} (k),
\]
where $\circledast$ is the convolution operator,  $\circ$ denotes the Hadamard or elementwise product, $K=\mathcal{F} k$ is called the convolution \emph{kernel}
and $k^\star$ the \emph{deapodization} factor.
We choose $K$ with finite width $k_w$, and the \emph{deapodization} factor can be   pre-computed as $k^\star=\{(\mathcal{F}^{\ast} K(\bm x))^{-1}$ if $u(\bm x)>0$, 0 otherwise\}. Several kernel functions have been proposed and employed in the literature, including truncated Gaussian, Kaiser-Bessel, 
or an interpolation kernel to minimize the worst-case
approximation error over all signals of unit norm \cite{fessler2003nonuniform}. 
    
\begin{figure}[tb]
    \centering
    \includegraphics[width=\textwidth]{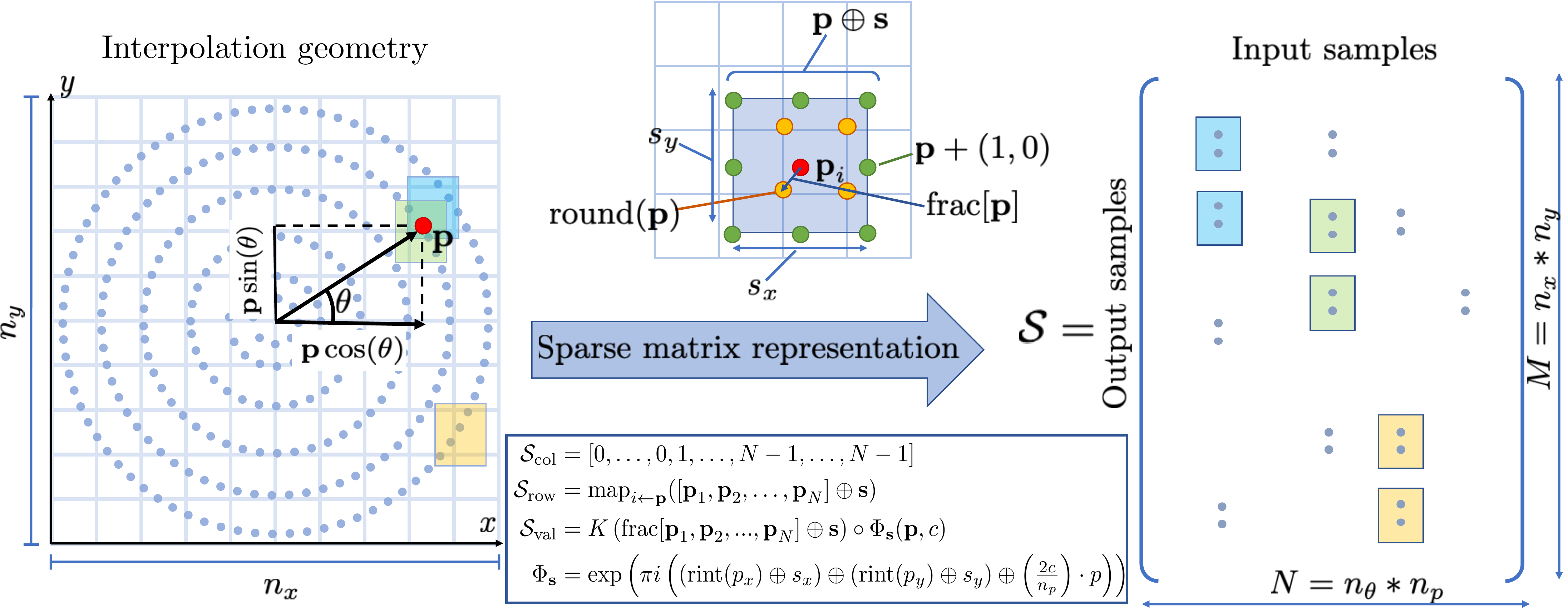}
    \caption{Sparse matrix $\mathcal{S}\in \mathbb{C}^{M\times N}$ (right) representation of a set of convolutional kernel windows of width $k_w=3$ with stencils $\bm s=\left(s_x,s_y\right)$, $s_x=(-1,0,1)$, $s_y=s_x^T$  (top),
    centered around a set of coordinates  $\bm{p}_i =p_i(\cos \theta_i,\sin\theta_i)+\tfrac{1}{2}(n_x, n_y)$   of each input point of the sinogram on the output image (left).
   \label{fig:sparse}
    \vspace{-0.4cm}}
    \end{figure}

The Fubini-Radon transform operator and its pseudo-inverse iRadon can be expressed using a sparse matrix (Fig. \ref{fig:sparse}) 
 to perform the interpolation~\cite{Ou:EECS-2017-90}:
\begin{align}
 \mathrm{Radon}(u(\bm x))&=\mathcal{F}_{p\leftarrow p^\prime}^{\dagger} \mathcal{S}^{\dagger}_{(\theta,p^\prime)\leftarrow(\bm p^\prime)} \mathcal{F}_{(\bm p^\prime)\leftarrow (\bm x)} k^\star(\bm x) \circ u(\bm x),\nonumber\\
\mathrm{iRadon}(\mathrm{Sino}(p,\theta))&=k^\star(\bm x) \circ  \mathcal{F}_{(\bm x)\leftarrow  (\bm p^\prime)}^{\dagger} \mathcal{S}_{(\bm p^\prime)\leftarrow(\theta,p^\prime)}\mathcal{D} \mathcal{F}_{p^\prime \leftarrow p} \mathrm{sino}(p,\theta),
\label{eq:fubini-radon}
\end{align}
where bold indicates 2D vectors such as $\bm p=(p_x,p_y)$, $\bm x=(x,y)$, and  $\mathcal{D}$ is a diagonal matrix to account for the density of sampling points in the polar grid. 


Fourier transforms and multiplication of $\mathcal{S}\in \mathbb{C}^{M\times N}$ with a sinogram in vector form $(1\times N$, $N=n_\theta \cdot n_p)$,  (sparse matrix-vector multiplication or SpMV) produces a tomogram of dimension $(1 \times M \rightarrow n_y \times n_x)$; multiplication with a stack of sinograms 
(sparse matrix-matrix multiplication or SpMM) produces the 3D $\mathrm{tomogram}(n_z, n_y, n_x))$. 
The diagonal matrix $\mathcal{D}$ can incorporate standard filters such as the Ram-Lak ramp, Shepp-Logan, Hamming, 
or a minimum residual filter based on the data itself \cite{mrfbp}. We can also employ the density filter solution that minimizes the difference with the impulse response (a constant $\bm 1$ in Fourier space) as 
$\arg \min_{\mathcal{D}_v} \|\mathcal{SD}_v-\bm{1}_{n_x \cdot n_y} \|$, with $\mathcal{D}_v$ as the vector of the diagonal elements of $\mathcal{D}$. Note that in this case, the matrix $\mathcal{D}=\mathrm{Diag}(\mathcal{D}_v )$ can be incorporated directly into $\mathcal{S}$ for better performance.

 The row indices and values of the sparse matrix are related to the coordinates where the kernel windows are added up on the output 2D image as
 $\bm{p}_i =p_i(\cos \theta_i,\sin\theta_i)+\tfrac{1}{2}(n_x, n_y)$,  
 with kernel window stencil $\bm s =(s_x,s_y)$, and $s_x=(-1,0,1)$, $s_y=s_x^T$ (for $k_w=3$) . The column index is simply given by the consecutive sequence of natural numbers $\mathbb{N}_1^N=[0,1,\dots, N-1]$, repeated $k_w^2$ times, and the row index and value are given by:
\begin{align*}
    \mathcal{S}_\mathrm{col}&=\mathbb{N}_0^{N-1}\otimes \mathbf{1}_{k_w^2}
    =[0,0,\dots,0,1,1,\cdots,N-1,\dots,N-1],\\
    \mathcal{S}_\mathrm{row} &=\lex([\bm p_1, \bm p_{2},...,\bm p_{N}]\oplus \bm s),\\
    \mathcal{S}_\mathrm{ val} &=K\left(\mathrm{frac}[\bm p_1, \bm p_{2},...,\bm p_{N}]\oplus \bm s\right)\circ \Phi_{\bm s}(\bm p,c),\\
    \Phi_{\bm s}(\bm p,c)&=\exp\left(
         \pi i \left((\mathrm{rint}(p_x)\oplus s_x)\oplus(\mathrm{rint}(p_y)\oplus s_y)\oplus \left( \tfrac{2 c}{n_p} \right) \cdot p\right)\right),
\end{align*}
where $\oplus \bm s$ is the broadcasting sum with the window stencil reshaped to dimensions $(1,1,k_w,k_w)$, $\lex(\bm p)= \mathrm{rint}(p_x)*n_y+\mathrm{rint}(p_y)$
is the lexicographical mapping from 2D to 1D index,
$K$ is the kernel function and $\mathrm{frac} [\bm p ]= \bm p -\mathrm{round}[\bm p ]$ is the decimal part, and $\otimes \bm 1$ represents the Kronecker product with the unit vector $\bm{1}_{(k_w)^2}=[1,1,\dots,1]$, for a window of width $k_w$ (see Fig. \ref{fig:sparse}).  $\cal S$ has at most $\text{nnz}=n_\theta \cdot n_p \cdot k_w^2$  non-zero elements, and the sparsity ratio is at most $ \tfrac{N\cdot k_w^2}{N\cdot M}=\tfrac{k_w^2}{n_x\cdot n_y}$, or $\tfrac{(k_w-1)^2}{n_x\cdot n_y}$ when the kernel is set to 0 at the borders.  We account for a possible shift $c$ of the rotation axis and avoid FFTshifts in the tomogram and radon spaces by applying a phase ramp as $\Phi_{\bm s}(\bm p,c)$, with $c=\tfrac{n_p}{2}$ when the projected rotation axis matches the central column of the detector. For better performance, 
 FFTshifts in Fourier space are incorporated in the sparse matrix by applying an FFTshift of the $ p$ coordinate, and by using a 
 $\bigl( \begin{smallmatrix}+1 & -1 \dots \\ -1 & +1 \dots \end{smallmatrix}\bigr)$  checkerboard pattern in the deapodization factor $k^\star$. 

\subsection{Parallel workflow}

The Fubini-Radon transform operates independently on each tomo and sinogram, so we can aggregate sinograms into chunks and distribute them over multiple processes operating in parallel. Denoising methods that operate across multiple slices can be handled using halos with negligible reduction in the final Signal-to-Noise-Ratio (SNR), while reducing or avoiding MPI neighborhood communication.

Pairs of sinograms are combined into a complex sinogram which is  processed simultaneously, by means of complex arithmetic operations, and is split back at the end of the reconstruction. We can limit the amount of chunks assigned to each process in order to avoid memory constraints. Then, when the data has more slices than what can be handled by all processes, it is divided up ensuring that each process operates on similar size chunks of data and all processes loop through the data. When the number of slices cannot be distributed equally to all processes, only the last loop chunk is split unequally with the last MPI ranks receiving one less slice than the first ones. 
\\
\indent The setup stage uses the experimental parameters of the data (number of pixels, slices and angles) and the choice of filters and kernels to compute the sparse matrix entries, deapodization factors and slice distribution across MPI ranks. During the setup stage, the output tomogram is initialized as either a memory mapped file, a shared memory window (whenever possible) or an array in rank-0 to gather all the results. 
\\
\indent Several matrix formats and conversion routines exist to perform the SpMV operation efficiently. In our implementation,  the sparse matrix entries  are first computed in Coordinate list (COO) format which contains a list of (row, column, value) tuples. Zero-valued and out-of-bound entries are removed, and then the sparse matrix is converted to  compressed sparse row (CSR) format, where the entries are sorted by column and row, and the row index is replaced by a compressed pointer. The sparse matrix and its transpose are stored separately and incorporate preconditioning filters and phase ramps to avoid all FFTshifts. The CSR matrix entries are saved in a cache file for reuse, with a hash function derived from the experimental parameters and filters to identify the corresponding sparse matrix from file. The FFT plans are computed at the first application and stored in memory until the reconstruction is restarted. When the data is loaded from file and/or the results are saved to disk, parallel processes pre-load the input to memory or flush the output from a double buffer  as the next section of the data is processed.
\\
\indent Our implementation  uses cuSPARSE and MKL libraries for the SpMV and FFT operations,  MPI for distributed parallelism through shared memory when available, or scatterv/gatherv and non-blocking double buffers for I/O and MPI operations. All these libraries are accessed  through Python, NumPy, CuPy, SciPy and mpi4py;  we also rely on h5py or tifffile modules to interface with data files. This framework also provides the capability to call TomoPy and ASTRA solvers on distributed architectures using MPI.
\\
\indent We used this framework to implement the most popular algorithms described in Sec. \ref{sec:irt}, namely FBP, SIRT, CGLS, and TV. To achieve high throughput, our implementation of  SIRT uses the Hamming preconditioning and the BB-step acceleration \cite{bb-step}, which provides 10-fold convergence rate speedup and makes it comparable to the conjugate gradient method but with fewer reductions and lower memory footprint. The CGLS implementation is based on the conjugate gradient squared method \cite{cgs}, and the TV denoising employs the split-Bregman \cite{goldstein2009split} technique.

\section{Experiments and Results}

The experimental evaluation presented herein is two-fold. We assess the performance of our implementation on both shared and distributed memory systems and on CPU and GPU architectures, and we also study how it compares to TomoPy, the state-of-the-art solution on X-ray sources, in terms of run time and quality of reconstruction.

We employ two different datasets for this analysis. The first one is a simulated Shepp-Logan phantom generated using TomoPy, with varying sizes to analyze the performance and scalability of the solution. The second one is an experimental dataset generated at  Lawrence Berkeley National Laboratory's Advanced Light Source during an outreach program with local schools out of a bread-crumb inserted at the  
micro-tomography beamline 8.3.2. The specifics of the experiments were: 25 keV X-rays, pixel size 0.65 microns, 200ms per image and 1313 angles over 180 degrees. The detector consisted of 20 micron LuAG:Ce scintillator and Optique Peter lens system with Olympus 10x lens, and PCO.edge sCMOS detector. The total experiment time, including camera readout/overhead, was around 6 minutes, generating a sinogram stack of dimension $(n_{z},n_{\theta},n_{p})$ = $(2160, 1313, 3620)$. 

We use two different systems for this evaluation. The first is the Cori supercomputer (\texttt{Cori.nersc.gov}), a Cray XC40 system comprised of 2,388 nodes containing two 2.3 GHz 16-core Intel Haswell processors and 128 GB DDR4 2133 MHz memory, and 9,688 nodes containing a single 68-core 1.4 GHz Intel Xeon Phi 7250 (Knights Landing) processor and 96 GB DDR4 2400 GHz memory. Cori also provides 18 GPU nodes, where each node contains two sockets of 20-core Intel Xeon Gold 6148 2.40 GHz, 384 GB DDR4 memory, 8 NVIDIA V100 GPUs (each with 16GB HBM2 memory). For our experiments, we use the Haswell processor and the GPU nodes.\footnote{Cori configuration page: \url{https://docs.nersc.gov/systems/cori/}}
The second system employed is CAM, a single node dual socket Intel Xeon CPU E5-2683 v4 @ 2.10GHz with 16 cores 32 threads each, 128 GB DDR4  and 4 NVIDIA K80 (dual GPU with 12 GB of GDDR5 memory each).
\begin{figure*}[ht!]
    \centering
    \includegraphics[width=0.39\linewidth]{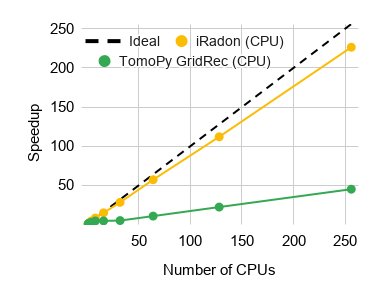}
    \includegraphics[width=0.39\linewidth]{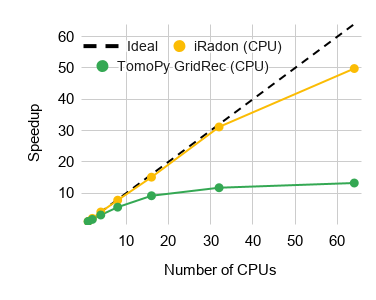}
    \includegraphics[width=0.2\linewidth]{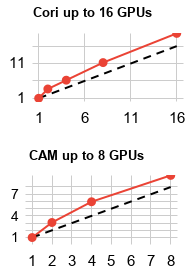}
    \caption{Speedup of iRadon and TomoPy-Gridrec algorithms on CPU Cori (left), CPU CAM (center) and GPU Cori and CAM (right) for a  $(n_{z},n_{\theta},n_{p})=(2048, 2048, 2048)$ simulation. The horizontal axis is the concurrency level and the vertical axis measures the speedup. 
    \label{fig:speedup_plots}}
\end{figure*}

The first experiment reports the performance results and scaling studies of our iRadon implementation and of TomoPy-Gridrec, when executed on both Cori and CAM, over the simulated dataset.
The primary objective is to compare their scalability using both CPUs and GPUs.
We executed both algorithms at varying levels of concurrency using a simulation size of $(2048, 2048, 2048)$.
On Cori, we used up to 8 Haswell nodes in a distributed fashion, only using physical cores in each node.
On CAM, we ran all the experiments on a single node, dual socket.
The speedup plots are shown in Fig.~\ref{fig:speedup_plots}.
The reported speedup is defined as $S(n,p) = \frac{T^*(n)}{T(n,p)}$ where $T(n,p)$ is the time it takes to run the parallel algorithm on $p$ processes with an input size of $n$, and $T^*(n)$ is the time for the best serial algorithm on the same input. 

\begin{figure*}[ht!]
    \centering
    \includegraphics[width=0.49\linewidth]{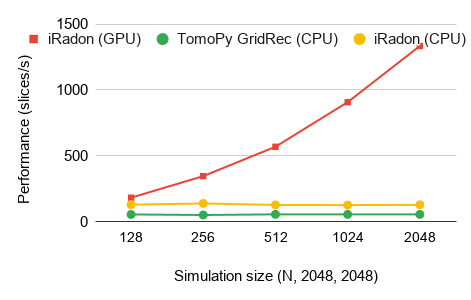}
    \includegraphics[width=0.49\linewidth]{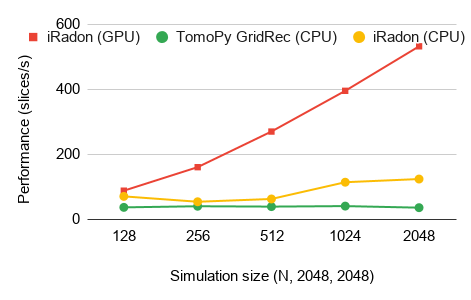} 
    \caption{Performance on Cori (left) and CAM (right), for varying sizes of simulated datasets as $(n_{z},n_{\theta},n_{p})=(N, 2048, 2048)$, running both the iRadon and TomoPy-Gridrec algorithms. The horizontal axis is the number of slices (sinograms) of the input data, and the vertical axis measures performance as slices reconstructed per second. CPU experiments employ 64 processes and GPU experiments use 8 on CAM and 16 on Cori.}
    \label{fig:runtime_plots}
\end{figure*}

First, we notice that the iRadon algorithm running on GPU has a super-linear speedup on both platforms.
This is unusual in general, however possible in some cases.
One known reason is the cache effect, i.e. the number of GPU changes, and so does the size of accumulated caches from different GPUs.
Specifically, in a multi-GPU implementation, super-linear speedup can happen due to configurable cache memory.
In the CPU case, we see a close to linear speedup.
On CAM, the performance decreases because of MPI oversubscribe, i.e. when the number of processes is higher than the actual number of processors available.

Finally, there is a clear difference in speedup results compared to the TomoPy-Gridrec implementation.
We believe that the main difference here is due to the fact that TomoPy only uses a multithreaded implementation with OpenMP, while our implementation relies on MPI.
For the purpose of comparison with our implementation, we use MPI to run TomoPy across nodes.

We also evaluate our implementation by running multiple simulations with a fixed number of angles and rays ($2048$) and varying number of slices ($128-2048$) on 64 CPUs and 8 GPUs.
Performance results in slices per second are shown in Fig.~\ref{fig:runtime_plots}.
One can notice that the GPU implementation of iRadon presents an increase in performance when the number of GPU increases.
This is a known behavior of GPU performance when the problem is too small compared to the capabilities of the GPU, and the device is not completely saturated with data, not taking full advantage of the parallelized computations.
For both platforms, our CPU implementation of iRadon performs significantly better than TomoPy.

In terms of raw execution time, TomoPy-Gridrec outperforms our iRadon implementation by a factor of $2.3$X when running on a single CPU on Cori.
On the other hand, the iRadon execution time using 256 CPU cores on Cori is $4.11$ seconds, outperforming TomoPy by a factor of $2.2$X.
Our iRadon version also ourperforms TomoPy by a factor of $1.9$X using 32 cores.
Our GPU implementation of iRadon runs in $1.55$ seconds using 16 V100 GPUs, which improves the CPU implementation (1 core) by a factor of $600$X, and runs $2.6$X faster compared with 256 CPU cores.
Finally, our GPU version of iRadon runs $7.5$X faster (using 2 GPUs) than TomoPy (using 32 CPUs), which could be considered the level of hardware resources accessible to average users.

\begin{figure}\CenterFloatBoxes
\begin{floatrow}
\ffigbox[\FBwidth]
{
\includegraphics[trim=00 27 00 15,clip,width =1.02 \linewidth]{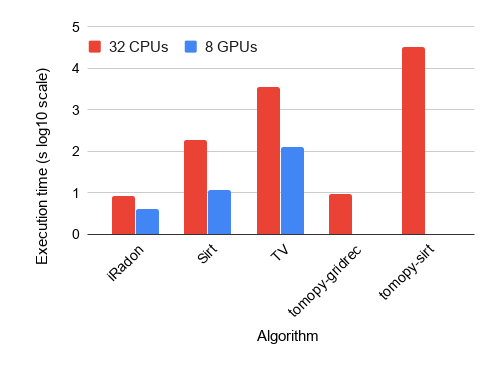}
}{\caption{Comparison of execution time (seconds in log10 scale) for different algorithms, reconstructing 128 slices of the bread-crumb dataset on CAM. SIRT and TV run for 10 iterations.
}\label{fig:bread_data}}
\hspace{0.5cm}
\ttabbox{
\begin{tabular}{c|c|c|c}
  \textbf{Alg.} & \textbf{CPU} & \textbf{GPU} & \textbf{SNR} \\ \midrule
  iRadon  & $\mathbf{0.14}$ & $\mathbf{0.07}$ & $3.51$ \\
  SIRT    & $3.13$ & $0.19$ & $17.11$ \\
  TV      & $57.8$ & $2.07$ & $\mathbf{17.78}$ \\
  \bottomrule
  \end{tabular}
}{\caption{Execution times for CPU and GPU (minutes) and SNR values for each reconstruction algorithm implemented. SNR is computed for a simulation of size (256, 1024, 1024).
}\label{tab:bread_runtimes}}
\end{floatrow}
\end{figure}

\vspace{-0.4cm}

The last experiment focuses on the analysis of the different algorithms implemented in this work, in terms of execution time and reconstruction quality.
Fig.~\ref{fig:bread_data} shows the reconstruction of $128$ slices of the bread-crumb experimental dataset on CAM (32 CPUs and 8 GPUs), for 3 different implemented algorithms: iRadon, SIRT, and TV, and also for TomoPy-Gridrec and TomoPy-SIRT. All iterative implementations (SIRT and TV) run for 10 iterations.
 Our iRadon implementation presents the best execution time for CPU ($9$ seconds), while on GPU, it runs in $4$ seconds. Our SIRT implementation outperforms TomoPy's by a factor of $175$X.
We report the SNR values (and corresponding execution times) of our implemented algorithms in Table~\ref{tab:bread_runtimes}, using a simulation dataset of size (256, 1024, 1024). We can observe how both SIRT and TV present the best results in terms of reconstruction quality.
\vspace{-0.2cm}

\begin{figure}
    \centering 
    \includegraphics[width=0.20\linewidth]{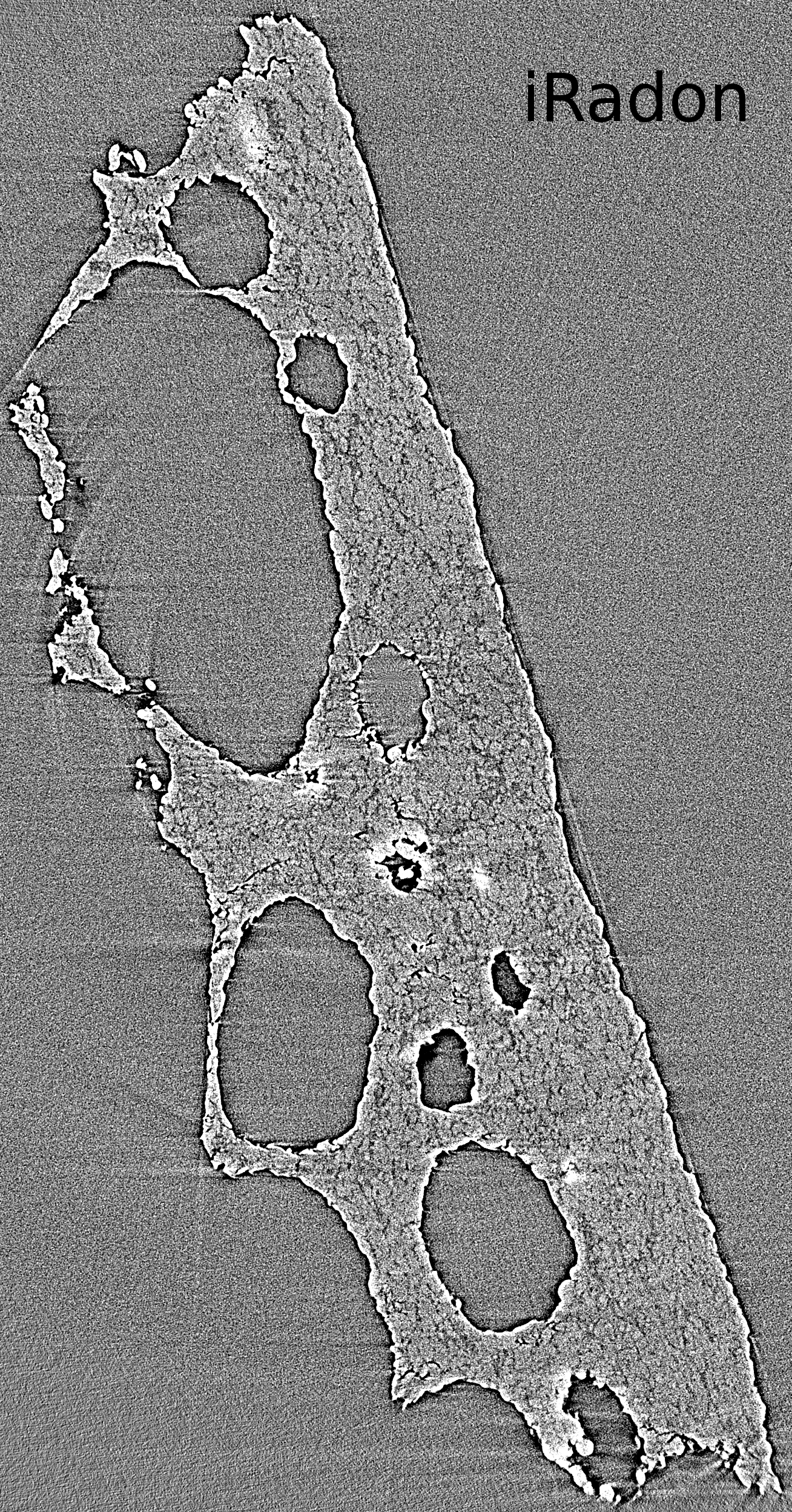}
    \hspace{0.05\linewidth}
    \includegraphics[width=0.20\linewidth]{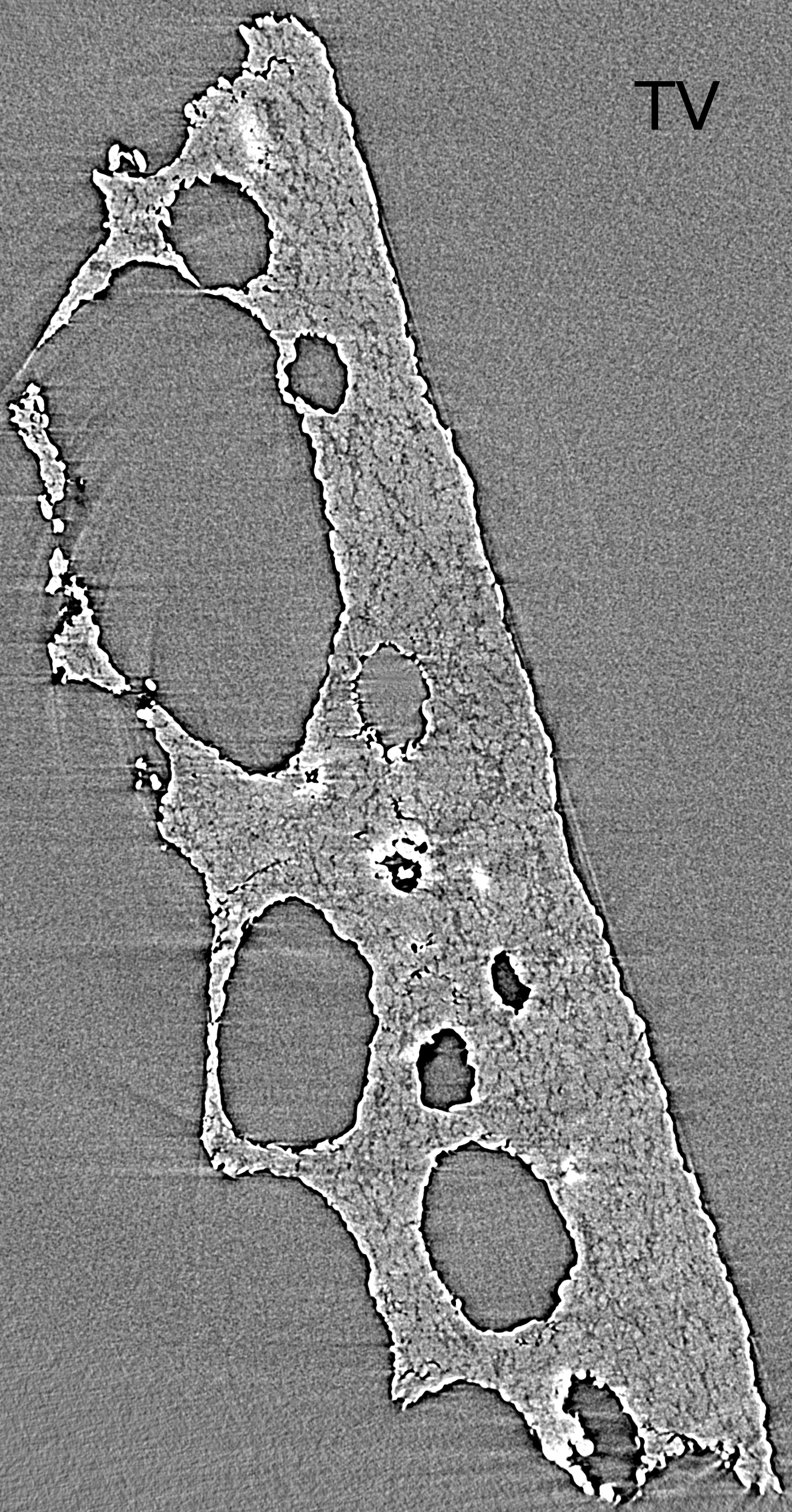}
        \hspace{0.05\linewidth}
    \includegraphics[width=0.19\linewidth]{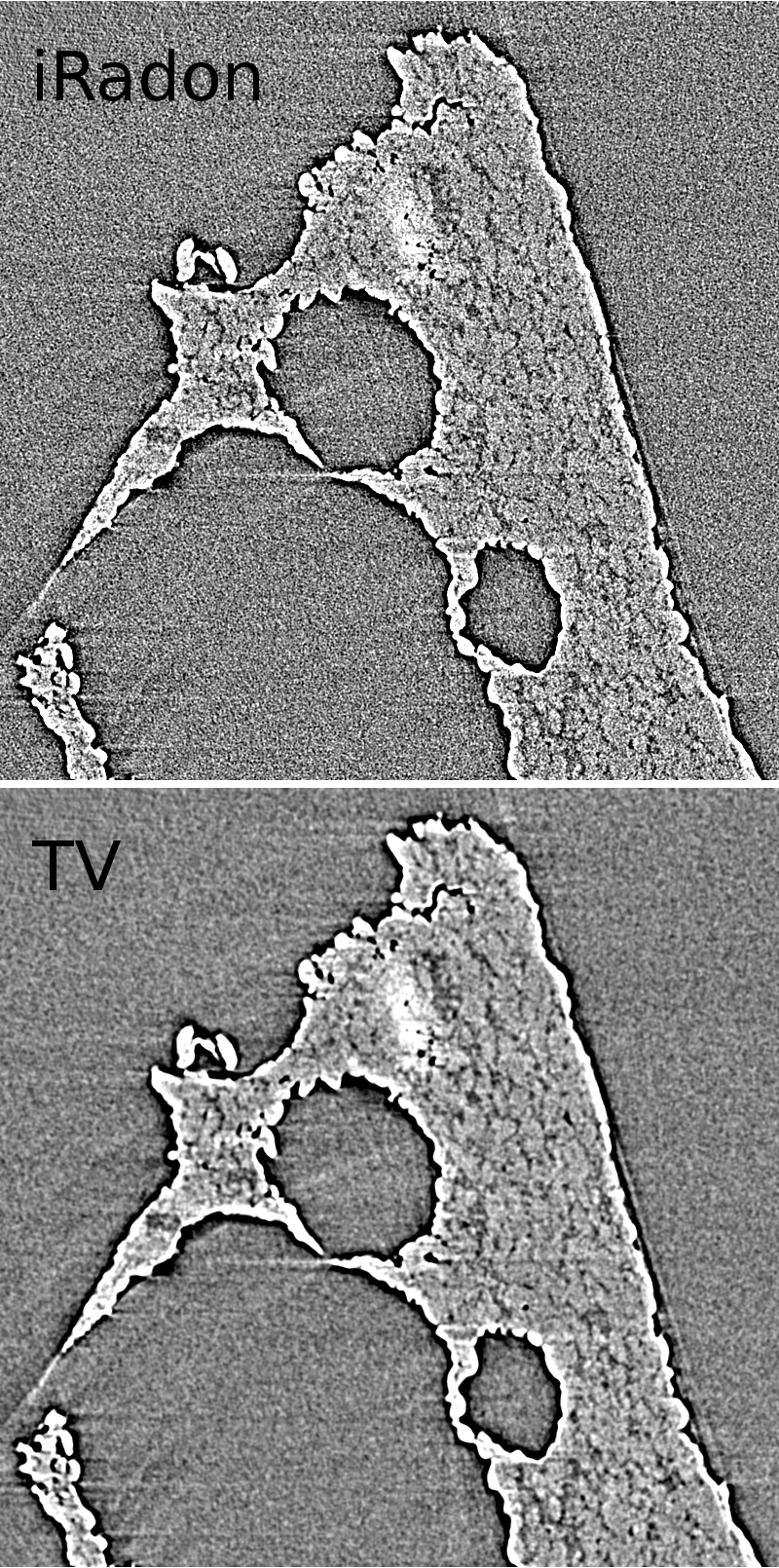}
    \caption{Example of reconstructed slice from the bread-crumb dataset using the iRadon and the TV algorithms. This visual result shows a better quality of reconstruction obtained using iRadon.}
    \label{fig:bread_rec}
\end{figure}

\vspace{-0.4cm}
Fig.~\ref{fig:bread_rec} shows a reconstructed slice of the bread-crumb data using the iRadon and the TV algorithms, along with a zoomed-in region of the same slice. The difference in reconstruction quality is minor in this case due to the dataset presenting high contrast and a large number of angles. Still, in the zoomed-out image we can appreciate higher contrast fine features on the TV reconstruction. Sparser datasets would be analyzed in the future to assess the performance of TV and iterative solutions on more challenging scenarios.

It is important to remark that all the execution times presented in this section refer to the solver portion of the calculations.
When running the TV algorithm on the complete bread-crumb data using 8 GPUs on CAM, for example, the solver time takes approximately $78\%$ of the total execution time ($44.82$ minutes). Most of the remaining time is taken by I/O ($18\%$) and gather ($2\%$).

\section{Conclusions}
\vspace{-0.2cm}
This paper presents a novel solution for tomography analysis based on fast SpMV operators. The proposed software is implemented in Python relying on CuPy, SciPy and MPI for high performance and flexible CPU and GPU reconstruction. As opposed to existing solutions, the software presented tackles the main requirements existing in tomography analysis: it can run over most hardware setups and can be easily adapted and extended into new solvers and techniques, while greatly simplifying deployment at new beamlines. The experimental results of this work demonstrate the remarkable performance of the solution, being able to iteratively reconstruct datasets of 68 GB in less than $5$ seconds using 256 cores and in less than $2$ seconds using 16 GPUs.
For the simulated datasets analyzed, the proposed software outperforms the reference tomography solution by a factor of up to $2.7$X, while running on CPU.
When reconstructing the experimental data, our implementation of the SIRT algorithm outperforms TomoPy by a factor of $175$X running on CPU.
The code of this project is also open source and available at \cite{xpack}.

As future work, we will employ CPU and GPU co-processing,  Block Compressed Row (BSR) format and sparse matrix-dense matrix multiplication (SpMM) to enhance the throughout of the solution. We will also explore the Toeplitz approach \cite{Ou:EECS-2017-90}, which permits combining the Radon transform with its adjoint into a single operation, while also avoiding the forward and backward 1D FFTs. Half-precision arithmetic is also probably sufficient to deal with experimental data with photon counting noise obtained with 16 bits detectors and can further improve performance by up to an order of magnitude using tensor cores. Generalization to cone-beam, fan beam or helical scan geometries using generalized Fourier slice methods \cite{cone-beam} will also be subject of future work. We will also explore the implementation of advanced denoising schemes based on wavelets 
or BM3D \cite{bm3d}, combining the operators presented in this work.

\section*{Acknowledgments}
\label{sect:Acknowledgements}
Work by S. M. was supported by Sigray, Inc. 
A. T. work was in part sponsored by Sustainable Research Pathways of the Sustainable Horizons Institute. P. E. was funded through the Center for Applied Mathematics for Energy Research Applications. T. P. is supported by the grant ``Scalable Data-Computing Convergence and Scientific Knowledge Discovery'', program
manager Dr. Laura Biven. D. P. is supported by the Advanced Light Source. This research used resources of the National Energy Research Scientific Computing Center (NERSC) and 
 the Advanced Light Source,  U.S. DOE Office of Science User Facility operated under Contract No. DE-AC02-05CH11231.

\bibliographystyle{splncs04}
\bibliography{bibliography}




\end{document}